\documentclass[aps,preprint,amsmath,amssymb]{revtex4}
\usepackage{color}
\usepackage{amsfonts}
\usepackage{graphicx}
\def\g5{\gamma_{5}}
\def\ga{\gamma}

\def\e{\epsilon}

\def\be{\begin{eqnarray}}
\def\ed{\end{eqnarray}}
\def \t{\tilde}
\def\non{\nonumber}
\def\l{\lambda}

\begin{document}
\title{\Large \bf  $D-\bar D$ mixing and rare $D$ decays
in the Littlest Higgs model with non-unitarity matrix}
\date{\today}

\author{ \bf  Chuan-Hung Chen$^{1,2}$\footnote{Email:
physchen@mail.ncku.edu.tw}, Chao-Qiang Geng$^{3,4}$\footnote{Email:
geng@phys.nthu.edu.tw} and Tzu-Chiang Yuan$^{3}$\footnote{Email:
tcyuan@phys.nthu.edu.tw}
 }

\affiliation{ $^{1}$Department of Physics, National Cheng-Kung
University, Tainan 701, Taiwan \\
$^{2}$National Center for Theoretical Sciences, Hsinchu 300, Taiwan\\
$^{3}$Department of Physics, National Tsing-Hua University, Hsinchu
300, Taiwan  \\
$^{4}$Theory Group, TRIUMF, 4004 Wesbrook Mall, Vancouver, B.C. V6T
2A3, Canada
 }

\begin{abstract}
We study the $D-\bar D$ mixing and rare $D$ decays in the Littlest
Higgs model. As the new weak singlet quark with the electric charge
of $2/3$ is introduced to cancel the quadratic divergence induced by
the top-quark, the standard unitary $3\times 3$
Cabibbo-Kobayashi-Maskawa matrix is extended to a non-unitary
$4\times 3$ matrix in the quark charged currents and $Z$-mediated
flavor changing neutral currents are generated at tree level. In
this model, we show that the $D-\bar D$ mixing parameter can be as
large as the current experimental value and the decay branching
ratio (BR)  of $D\to X_u \ga$ is small but its direct CP asymmetry
could be $O(10\%)$. In addition, we find that the BRs of $D\to X_u
\ell^{+} \ell^{-}$, $D\to X_u\nu \bar \nu$ and $D\to \mu^{+}
\mu^{-}$ could be enhanced to be $O(10^{-9})$, $O(10^{-8})$ and
$O(10^{-9})$, respectively.

\end{abstract}
\maketitle

\section{ Introduction}

As the observation of the $B_s-\bar B_s$ mixing in 2006 by CDF \cite{CDF},
 all neutral pseudoscalar-antipseudoscalar oscillations ($P-\bar P$)
in the down type quark systems have been seen.
In the standard  model (SM),
the most impressive features of flavor physics are the
Glashow-Iliopoulos-Maiani (GIM) mechanism \cite{GIM} and the large
top quark mass.
The former results in the cancellation between the lowest order
short-distance (SD) contributions of the first two generations to
the mass difference $\Delta m_K$ in the $K^{0}$ system,
while
the latter makes $\Delta m_{B_q}$ $(q=d, s)$ in the $B_{q}$ systems dominated by the SD effects \cite{Hagelin}.
In addition, these
features also lead to sizable
flavor changing neutral
currents (FCNCs) from box and penguin diagrams, which contribute to the rare decays, such as
$K\to \pi\nu\bar{\nu}$ and $B\to K^{(*)}\ell\bar{\ell}$.
It is known
that these processes could be good candidates to probe
 new physics effects \cite{new_KB1,new_KB2,new_KB3}.
 However, it is clear that
 the new physics signals deviated from the SM
predications for the $P-\bar P$ mixings and rare FCNC decays have to
wait for precision measurements on these processes.

Unlike $K$ and $B_q$ systems,  the SD
contributions to charmed-meson FCNC processes, such as the $D-\bar D$
mixing \cite{DD_SM} and the decays of
 $c\to u \ell^{+} \ell^{-}$ and $D\to \ell^{+}
\ell^{-}$ \cite{BGHP_PRD52}, are highly suppressed
due to the stronger GIM mechanism and weaker
heavy quark mass enhancements in the loops.
On the other hand,
it is often claimed that the long-distance (LD) effect
for the $D-\bar D$ mixing
should be the dominant contribution in the SM.
Nevertheless, because the nonperturbative hadronic
effects are hard to control, the result
is still inconclusive
\cite{order3,Petrov,order2-1,order2-2}.
Recently, BABAR \cite{babar_D} and BELLE \cite{belle_D,Staric} collaborations have reported the evidence for the $D-\bar D$ mixing with
 \be
 \label{D1}
 x'^{2}&=&(-0.22\pm0.30\pm0.21)\times 10^{-3}\,,\non\\
  y'&=& (9.7 \pm 4.4  \pm 3.1 ) \times
  10^{-3}\,,
  %\ \ \ [{\rm BABAR}]
\ed
and
\be
\label{D2}
x&\equiv& {\Delta m_{D}\over \Gamma_{D}}\;=\; (0.80\pm0.29\pm0.17)\%\,,\non\\
y&\equiv& {\Delta \Gamma_{D}\over 2\Gamma_{D}}\;=\;(0.33\pm0.24\pm0.15)\%\,,\non\\
  y_{CP}&=&(1.31\pm 0.32 \pm 0.25)\% \,,
  %\ \ \ [{\rm BELLE}]
 \ed
respectively,
where $x'=x\cos\delta+y\sin\delta$ and $y'=-x\sin\delta+y\cos\delta$ with the assumption of CP conservation and $\delta$ being the relative
 strong phase between the amplitudes for the doubly-Cabbibo-suppressed
 $D\to K^{+}\pi^{-}$ and
 Cabbibo-favored $D\to K^{-}\pi^{+}$ decays \cite{PDG06,Nir}
and $y_{CP}=\tau(D\to K^{-}\pi^{+})/\tau(D\to K^{+}K^{-})-1$.
Moreover, no evidence for CP violation is found.
The combined results of Eqs. (\ref{D1}) and (\ref{D2}) at the $68\%$ C.L.
are \cite{Combined}
\be
\label{D12}
x&=& (5.5\pm 2.2)\times 10^{-3}\,,\non\\
y&=& (5.4\pm2.0)\times 10^{-3}\,,\non\\
\delta&=&(-38\pm46)^{0}\,.
\ed
Note that the upper bound of $x<0.015$ at $95\%$C.L. can be extracted
from the BELLE data in Eq. (\ref{D2}) \cite{belle_D,Staric}.
The evidences of the mixing parameters by
BABAR and BELLE collaborations
reveal that the
era of the rare charmed physics has arrived.
The results in Eq. (\ref{D12}) can
 not only  test the SU(3) breaking effects
 for the $D-\bar D$ mixing
\cite{order2-2,Petrov}, but also  examine  new physics beyond the SM
\cite{Nir,Combined,GPP,New_D,Blanke}.

It is known that a
 straightforward way to enhance the rare $D$ processes is to
include some new heavy quarks within the framework of the SM. For
instance, if a new heavy quark with the electric charge of $-1/3$ is
introduced, it could affect the $D$ system since the extra down type quark
violates the GIM mechanism.
However, the constraint on this heavy quark is quite strong
as it could also lead to
FCNCs for the down type quark sector at tree level,
which are strictly limited by
the well measured rare $K$ and $B$ decays,
such as $K_L\to \mu^{+} \mu^{-}$ and $B\to
X_s\gamma$ \cite{CCK_PRD61}.
On the other hand, if the charge
of the new heavy quark is
 $2/3$, it could generate interesting tree level FCNCs for the
up type quark sector, for which the constraints are much weaker.
In this paper, we will study  $D$ physics based on
 a new weak singlet upper quark.

It has been known that in the framework of the Littlest Higgs model
\cite{LH}, there exists a new $SU(2)_L$ singlet vector-like up quark
\cite{HLMW}, hereafter denoted by $T$.  Since the number of down
type quarks is the same as that in the SM,
the standard unitary $3\times 3$
Cabibbo-Kobayashi-Maskawa (CKM) matrix is extended to a non-unitary
$4\times 3$ matrix in the charged currents.
Moreover,
$Z$-mediated FCNCs for the up quark sector
are generated at tree level.
%We note that in
In Ref.~\cite{Lee}, it has been shown that the contributions
of this new quark to the rare $D$ processes are small and
cannot reach the sensitivities of future experiments
\cite{Lee,FP_PRD73}.
%However, in
In this paper, we will demonstrate that
by adopting some plausible scenario, the effects could not only
generate a large $D-\bar D$ oscillation but also marginally reach
the sensitivity proposed by BESIII for the rare $D$ decays \cite{bes3}.
%%%%%
%%%%%
We note that the implication of the new data on the $D-\bar D$
mixing in the Littlest Higgs model with T-parity has been recently
studied in Ref. \cite{Blanke}.
%%%%%%%

 The paper is organized as follows. In Sec.~\ref{sec:non-unitary}, we
investigate that when a gauge singlet $T$-quark is introduced in the
Littlest Higgs model, how the non-unitary matrix for
the charged current and the tree level $Z$-mediated FCNC are formed.
By using the leading perturbation, the mixing matrix elements related
to the new parameters in the Littlest Higgs model are derived. In
addition, we study how to get the  small mixing matrix element for
$V_{u(c)b}$, which describes the $b\to u(c)$ decays. In
Sec.~\ref{sec:rareD}, we discuss the implications of the non-unitarity on
the $D-\bar D$ mixing and rare $D$ decays by presenting some numerical analysis.
Finally, we summarize our results in Sec.~\ref{sec:Concl}.

\section{Non-unitary mixing matrix in the Littlest
model}\label{sec:non-unitary}

To study the new flavor changing effects in the Littlest Higgs model, we
start by writing the Yukawa interactions for the up quarks
to be \cite {HLMW,Lee}
 \be
  {\cal L}_{Y}&=& \frac{1}{2} \l_{ab} f \e_{ijk} \e_{xy} \chi_{ai}
  \Sigma_{jx} \Sigma_{ky} u^{\prime c}_{b} + \l_0 f T T^{c} +h.c.\,,
 \ed
where $\chi^{T}_{1}=(d_1, u_1, 0)$, $\chi^{T}_{2}=(s_2, c_2, 0)$,
$\chi^{T}_3=(b_3, t_3, T)$, $u^{\prime}_{b}$ is the weak singlet
 and $\Sigma=e^{i\Pi/f} \Sigma_0 e^{i\Pi^{T}/f}$
with
 \be
 \Sigma_0=\left(\begin{array}{ccc}
 & & { \openone}_{2\times2} \\
 & 1 &  \\
{\openone }_{2\times2} & & \end{array}\right)\,,\ \ \
      \Pi=\left( \begin{array}{ccc}
 & h^\dagger/\sqrt{2} & \phi^\dagger \\
h/\sqrt{2} & & h^\ast/\sqrt{2} \\
\phi & h^T/\sqrt{2} & \end{array} \right)\,.
 \ed
The scale $f$ denotes the global symmetry spontaneously breaking
scale, which,  as usual, could be around $1$ TeV.
Consequently, the $4\times 4$
up-quark mass matrix is given by \cite{Lee}
 \be
 M_{U}&=& \left[
            \begin{array}{ccc}
              \left(
                \begin{array}{c}
                  i\l_{ij}v \\
                \end{array}
              \right)_{3\times 3} & | & 0 \\
             -\ -\ - & - &\  - \\
                              0\ \ 0 \ \ \l_{33} f & |& \l_0 f \\
            \end{array}
          \right]\,.
 \ed
We remark that the quadratic divergences for the Higgs mass from
one-loop diagrams involving $t$ and $T$ get exactly cancelled as
shown in Ref. \cite{Perelstein}. Moreover, for other quarks other
than the top quark, the one-loop quadratic divergent contributions
do not necessitate fine-tuning the Higgs potential as the cutoff is
around 10 TeV for $f\sim 1$ TeV due to the small corresponding
Yakawa couplings. That is why there is no need to introduce extra
singlet states $T$ \cite{Perelstein,Schmaltz}.

To obtain the quark mass hierarchy of $m_t\gg
m_c \gg m_u$,
we can choose a basis such that the up-quark mass matrix
is \cite{QM}
 \be
M_{U}=\left(
  \begin{array}{cc}
      {\bf \hat m_{U}} &  0\\
         {\bf h} f & \l_0 f  \\
  \end{array}
\right) \label{eq:mass}
 \ed
where $\hat m_{U ij}=\delta_{ij} \l_i v/\sqrt{2} \equiv m_{i}$ is
diagonal matrix and ${\bf h}=(h_1,\,h_2,\,h_3)$.  The $h_i$ is
related to $\l_{33} $ by $h_i=\t V^{U_R}_{i3} \l_{33} $ and ${\bf h
h^{\dagger}}= |\l_{33}|^2$, in which $\t V^{U_R}$ is the unitary
transformation for the right-handed up quarks. We note that $m_i$ are
not the physical masses and  in principle their magnitudes could be
as large as the weak scale. In order to preserve the hierarchy in the
quark masses, one expects that $m_3> m_2 > m_1$.
Furthermore, in terms of this basis, the charged and neutral currents, defined by
 \be
  {\cal L}^{C}=\frac{g}{\sqrt{2}} J^{-}_{\mu} W^{+ \mu}-
  \frac{g}{\sqrt{2}\tan\theta } J^{-}_{\mu} W^{+\mu}_{H} + h.c. \,,
  \non\\
 {\cal L}^{N}=\frac{g}{\cos\theta_W}
  \left( J^{\mu}_{3} - \sin^2\theta_W J^{\mu}_{em}\right) Z_{ \mu}+
  \frac{g}{\tan\theta} J^{\mu}_{3} Z_{H\mu} + h.c. \,,
 \ed
are expressed by
 \be
 J^{-}_{\mu}&=& \bar U_L \ga_{\mu}  \tilde V^{0} a_{V}  D_L \,, \non\\
 J^{\mu}_{3}&=& \frac{1}{2} \bar U_L \ga^{\mu}  \tilde V^{0}a_{V} \tilde V^{0\dagger}
 U_L -\frac{1}{2} \bar D_{L} \ga^{\mu} D_{L}\,,
 \label{eq:cn0}
 \ed
respectively, where $U^{T}=(u_1,\,c_2,\,t_3,\,T)$, $D^{T}=(d,\,s,\,b)$,
$a_{V}=\rm diag(1,1,1,0)$ and
 \be
 \tilde V^{0}&=& \left(
                   \begin{array}{cc}
                     \left(V^{0}_{CKM}\right)_{3\times 3}\ \  &  0 \\

                     0\ \  & 1 \\
                   \end{array}
                 \right)
 \ed
with $V^{0}_{CKM} V^{0\dagger}_{CKM}=\openone_{3\times 3}$. The null
entry in $a_{V}$ denotes the new $T$-quark being a weak singlet;
and without the new $T$-quark, $V^{0}_{CKM}$ is just the CKM matrix.
Since the down quark sector is the same as that in the SM,
we have set the unitary
transformation $U^{D_L}$ to be an identity  matrix.

For getting the physical eigenstates, the mass matrix in
Eq.~(\ref{eq:mass}) can be diagonalized by unitary matrices
$V^{U_{L, R}}$ so that we have
 \be
 M^{\rm diag}_{U} M^{\dagger \rm diag}_{U}&=& V^{U_L} M_{U}
 M^{\dagger}_{U} V^{U_L^\dagger} \label{eq:dia}
 \ed
and
  \be M_{U}M^{\dagger}_{U}=\left(
  \begin{array}{cc}
      {\bf \hat m_{U} \hat m_{U}^{\dagger}} &  {\bf \hat m_{U} h^{\dagger}} f\\
         {\bf h \hat m^{\dagger}_{U}} f & (|\l_{33}|^2+ |\l_0|^2 ) f^2  \\
  \end{array}
\right)\,. \label{eq:mass2}
 \ed
Since $(|\l_{33}|^2+ |\l_0|^2 ) f^2 $ is much larger than other
elements, we can take the leading order of the perturbation in $h_i
m_{i}/f$ $(i=1, 2, 3)$. According to Eq.~(\ref{eq:dia}), the leading
expansion is given by
 \be
  M^{\rm diag}_{U} M^{\dagger \rm diag}_{U}&=& V^{U_L} M_{U}
 M^{\dagger}_{U} V^{U_L^\dagger} \approx (1+\Delta_L)M_{U}
 M^{\dagger}_{U} (1 -\Delta_L)\,. \label{eq:massdia}
 \ed
By looking at the off-diagonal terms $(M^{\rm diag}_{U} M^{\dagger \rm
diag}_{U})_{i4(4i)}$, we can easily get
 \be
\Delta_{Li4}\approx  -\Delta_{L4i}= - \frac{h_i m_i f}{(
|\l_{33}|^2+ |\l_0|^2 ) f^2-m^2_{i} } \label{eq:deltai4}
 \ed
with $i\neq 4$. From the diagonal entries, if we set the light
quark masses to be $m_u\approx m_c\approx 0$, we obtain
 \be
   0 \approx m^{2}_{u_j}&\approx & m^2_{j} + 2\Delta _{Lj4}
  (M_{U}M^{\dagger}_{U})_{4j}\,,\non \\
  \Delta _{Lj4}&\approx &-\frac{1}{2} \frac{m_j}{h_j f}
  \label{eq:vj4}
 \ed
with $j=1,\, 2$. To be consistent with Eq.~(\ref{eq:deltai4}), at
the leading expansion the relation
%$2h^2_{j}=(|\l_{33}|^2+|\l_0|^2)$
\be
2h^2_{j}&=&(|\l_{33}|^2+|\l_0|^2)
\label{eq:ft}
\ed
should be satisfied.
%%%%%%%%
%%%%%%%%
We emphasize that the choice of Eq. (\ref{eq:ft}) is somewhat
fine-tuned in order
to have Eqs. (\ref{eq:deltai4}) and (\ref{eq:vj4}) simultaneously.
%%%%%%%%
Since the top-quark is much heavier than other
ordinary quarks,  we  have $2h^2_{3}\approx
(1-m^2_t/m^2_3)(|\l_{33}|^2+|\l_0|^2)$ if $f>  m_3> m_t$.
Similarly, one obtains the flavor mixing effects for $i\neq j \neq
4$ to be
 \be
 \Delta_{Lij}&=& \frac{h_i h_j m_{i}m_{j}}{m^2_i-m^2_j}
 \frac{f^2 [2(|\l_{33}|^2+|\l_0|^2)f^2-(m^2_i+m^2_j)]}{\left( (|\l_{33}|^2+|\l_0|^2)f^2-m^2_j \right)
 \left((|\l_{33}|^2+|\l_0|^2)f^2-m^2_i \right)}\,.
 \label{eq:deltaij}
 \ed
%M
After diagonalization,
the currents become
 \be
 J^{-}_{\mu}&=& \bar U_L \ga_{\mu} V^{U_L} \tilde V^{0} a_{V}  D_L = \bar U_L \ga_{\mu} V^{U_L} V^{0}   D_L\,, \non\\
 J^{\mu}_{3}&=& \bar U_L \ga^{\mu} V^{U_L} \tilde V^{0}a_{V} \tilde V^{0\dagger}V^{\dagger U_L}
 U_L=\bar U_L \ga^{\mu} V^{U_L} V^{0} V^{0\dagger} V^{\dagger U_L} U_L\,,
 \label{eq:cn}
 \ed
where $U^{T}=(u,\,c,\,t,\,T)$,
 \be
  V^{0}= \tilde V^{0} a_{V}= \left(
                   \begin{array}{cc}
                     \left(V^{0}_{CKM}\right)_{3\times 3}\ \  &  0 \\

                     0\ \  & 0 \\
                   \end{array}
                 \right)
                 \ed
and diag$(V^{0} V^{0\dagger})=a_{V}$.
Since the 4-th component of $a_V$ is
different from the first 3 ones, it is obvious that the matrix
$V\equiv V^{U_L} V^{0}$ associated with the charged current does not
satisfy unitarity. In addition, $V^{U_L} a_{V} V^{ U_L^\dagger}$, which
is associated with the neutral current, is not the identity matrix.
As a
result, $Z$-mediated FCNCs at tree
level are induced. According to Eq.~(\ref{eq:cn}), we see that
 \be
 VV^{\dagger}=V^{U_L}a_{V}V^{ U_L^\dagger}
 \ed
 which is just the same as the effects of the $Z$-mediated FCNCs.
 Due to  $V$ being a non-unitary
 matrix, one finds
 \be
   (VV^{\dagger})_{ij}= V_{i4} V^{*}_{j4} \,. \label{eq:nonunitary}
 \ed
Consequently, the interesting phenomena arising from non-unitary
matrix elements are always related to $V_{i4} V^{*}_{j4}=\Delta_{L
i4} \Delta_{j4}$.
We note that as we do not particularly address CP problem, in
most cases, we set the parameters to be real numbers.

It has been known that enormous data give strict bounds on the
flavor changing effects. In particular, the pattern
describing the
charged current has been fixed
quite well. Any new parametrization should respect these
constraints. It should be interesting to see the relationship with
and without the new vector-like $T$-quark. From Eq.~(\ref{eq:cn}), we
know that the new flavor mixing matrix for the charged current is
given by $V=V^{U_L} V^0$.  At the leading order perturbation, one
gets
 \be
 V= V^{U_L} V^0 \approx (1+\Delta_{L}) V^0 =V^{0} + \Delta_L V^{0}\,.
 \ed
If $V^{0}_{tb} \sim 1$ is taken, one finds that $V_{ub}\approx
V^{0}_{ub} + \Delta_{L 13}$ and $
 V_{cb}  \approx  V^{0}_{cb} + \Delta_{L23}$.
In terms of Eq.~(\ref{eq:deltaij}) and $h_1=h_2 \approx h_3$, the
relations $\Delta_{L13}\approx  - m_{1}/m_3$ and
$\Delta_{L23}\approx  - m_{2}/m_3$ are obtained. Hence, in our
approach, we have
 \be
 V_{us}\approx V^{0}_{us}-\frac{m_1}{m_2}\,,\ \ \  V_{ub} \approx V^{0}_{ub} - \frac{m_1}{m_3}\,, \ \ \ V_{cb} \approx V^{0}_{cb} -
  \frac{m_2}{m_3}\,.
 \ed
 From these results, it is clear that when  the $T$-quark decouples from ordinary quarks,
$V_{us}\to V^{0}_{us}$, $V_{ub}\to V^{0}_{ub}$ and $V_{cb}\to
V^{0}_{cb} $, while $m_1/m_2\to m_u/m_c$, $m_1/m_3\to m_u/m_t$ and
$m_2/m_3 \to m_c/m_t$, respectively. According to the observations
in the decays of $b\to u \ell \bar\nu_{\ell}$ and $b\to c \ell
\bar\nu_{\ell}$, the corresponding values have been determined to be
$|V_{ub}|=3.96\pm 0.09\times 10^{-3}$ and
$|V_{cb}|=42.21^{+0.10}_{-0.80}\times 10^{-3}$, respectively
\cite{PDG06}. Since $V^{0}_{ij}$ and $m_i$ are free parameters, to
satisfy the experimental limits
with interesting phenomena in
low energy physics, it is reasonable to set the orders of magnitude
for $m_{1}/m_3$ and $m_2/m_3\, (m_1/m_2)$ to be $O(10^{-2})$ and
$O(10^{-1})$, respectively. Consequently, the non-unitary effects on
the rare charmed meson decays governed by $V_{14} V^{*}_{24}$ could be as
large as $\Delta_{14} \Delta_{24}\sim O(10^{-4})$, which
could
be one order of magnitude larger than those in Ref.~\cite{Lee}.

\section{$D-\bar D$ mixing and rare $D$ decays}
\label{sec:rareD}

\subsection{$D-\bar D$ mixing}

It is well known that the GIM mechanism has played an important role in
the $K-\bar K$ oscillation in the SM.
In addition,  due to the top-quark
in the box and penguin diagrams,
$B_q-\bar B_q$ mixings are dominated by the SD effects,
which are consistent with the data \cite{PDG06}.
 On the contrary,  for the $D-\bar D$ mixing
the GIM cancellation further suppresses the mixing effect to be
$\Delta m_D \sim O(m^4_s/m^2_W m^2_c)$ \cite{DD_SM}
and the bottom
quark contribution
actually is a subleading
effect due to the
suppression of $(V_{ub}
V^*_{cb})^2$. In the SM, the SD contribution to the mixing parameter
is $O(10^{-7})$ \cite{GP_PLB625}.
However, the LD contribution to
 the  mixing is believed to be dominant.
Due to the nonperturbative hadronic
effects,
the result
is still uncertain with the prediction on the mixing parameter ranging  from $O(10^{-3})$
\cite{order3} to $O(10^{-2})$ \cite{Petrov,order2-1,order2-2}.
Nonetheless, the mixing parameters shown
in Eq. (\ref{D12}) could arise from the LD contribution.
Thus, it is important to have a better understanding of
 the LD effect.
On the other hand,
it is also possible that the mixings in Eq. (\ref{D12}) could result from
new physics.
In the following, we will concentrate on the Littlest Higgs model.

\begin{figure}[htbp]
\includegraphics*[width=2.5 in]{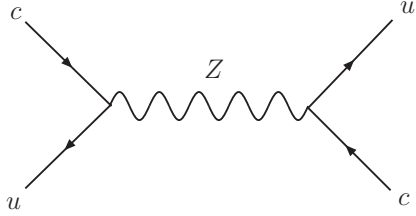}
\caption{$Z$-mediated flavor diagram with $|\Delta C|=2$.}
 \label{fig:ZDD}
\end{figure}

In the quark sector of the Littlest Higgs model due to the
introduction of a new weak singlet,  a direct impact on the low
energy physics is the FCNCs at tree level. According to
Eq.~(\ref{eq:cn}), the most attractive process with $|\Delta C|=2$ via
the $Z$-mediated $c-u-Z$ interaction, illustrated in Fig.~\ref{fig:ZDD},
is given by
 \be
  {\cal H}(|\Delta C|=2) &=& \frac{g^2}{4m^2_{W}}
    \left( V_{14}
  V^{*}_{24}\right)^2
  \bar u \ga_{\mu} P_L c \, \bar u \ga^{\mu} P_{L} c \,, \non \\
  &=& \frac{2 G_F}{\sqrt{2}}
  \left( V_{14}
  V^{*}_{24}\right)^2
  \bar u \ga_{\mu} P_L c \, \bar u \ga^{\mu} P_{L} c \,.
 \ed
%%%%%%%%%%%%%%%%%%%%%%%%%%%%%%%%%%%%%%%%%%%%%%%%%%%%%%%%%%%%%%%%%%%%%%%%%
%%%%%%%%%%%%%%%%%%%%%%%%%%%%%%%%%%%%%%%%%%%%%%%%%%%%%%%%%%%%%%%%%%%%%%%%%
In terms of the hadronic matrix element, defined by
 \be
  \langle \bar D | (\bar u c)_{V-A} \, (\bar u  c)_{V-A}|
 D\rangle=\frac{8}{3} B_{D} f^{2}_{D} m^2_{D}\,,
 \ed
the mass difference for the $D$ meson is \cite{Lee}
 \be
  \Delta m_{D} &\approx & \frac{\sqrt{2}}{3}G_{F} f^2_D m_D B_{D} |(V_{14}
 V^{*}_{24})^2| \,.\label{eq:md}
 \ed
If we assume no cancellation between new physics and SM contributions,
 by taking
$\tau_D=1/\Gamma_{D}=6.232\times 10^{11}$ GeV$^{-1}$,
$f_{D}\sqrt{B_{D}}=200$ MeV \cite{BD,Lattice} and $m_{D}=1.86$ GeV
and using Eq. (\ref{D12}),
 we obtain
 \be
  \zeta_{0}\equiv |V_{14}V^*_{24}|=|\Delta_{L14} \Delta_{L24}|&=& (1.47\pm 0.29)
   \times
  10^{-4}\,, \label{eq:limit}
 \ed
which is in the desirable range. In other words,
the result in Eq. (\ref{eq:limit})  demonstrates that
the non-unitarity in the Littlest Higgs model could enhance
the $D-\bar D$ mixing at the observed level.
%,
We note that the limit of $x<0.015$ ($95\%$C.L.) leads to
\be
\label{NLimit}
\zeta_{0}&<&2.5\times 10^{-4}.
\ed
In addition, we note that cancellation between the LD effect in the SM and the SD one from new physics could happen.
In this case,
the  values in
Eqs.~(\ref{eq:limit}) and (\ref{NLimit}) could be relaxed.

\subsection{$D\to X_u\ga$ decay}

In the SM, the $D$-meson FCNC related processes are all
suppressed  since the internal fermions in the loops are all much
lighter than $m_{W}$. For the decay of $D\to X_u \ga$,
without QCD corrections, the branching ratio is
$O(10^{-17})$; and  it becomes $O(10^{-12})$ when one-loop QCD
corrections are included \cite{BGHP_PRD52}. However, it is found
that the two-loop QCD corrections can boost the BR to be as large as
$3.5\times 10^{-8}$ \cite{GHMW}. It should be interesting to see how
large the non-unitarity effect on $c\to u\ga$ is in the Littlest
Higgs model.

\begin{figure}[htbp]
\includegraphics*[width=6.5in]{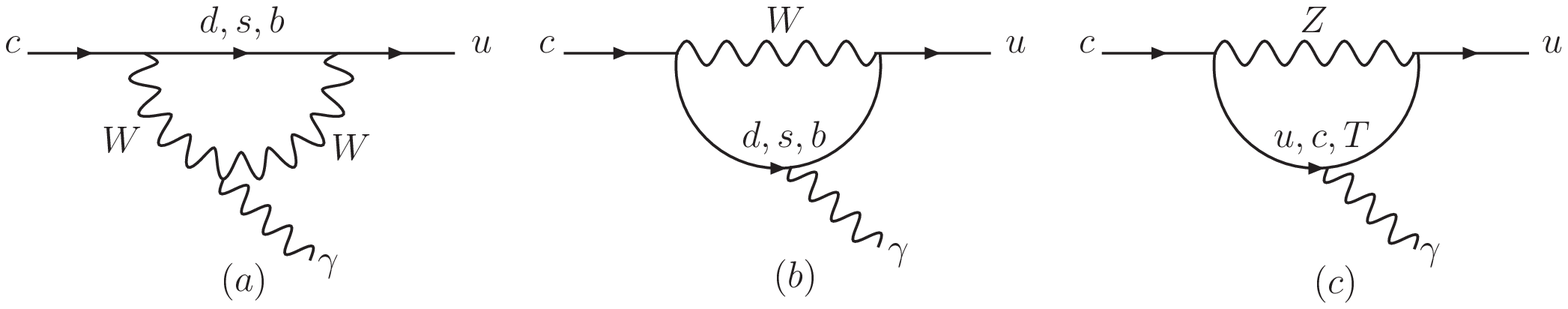}
\caption{Flavor diagrams for $c\to  u \ga$.}
 \label{fig:WZ}
\end{figure}

To study the radiative decay of $c\to u \ga$, we write the effective Lagrangian to be
 \be
 {\cal L}_{c\to u \ga}= -\frac{ G_F}{\sqrt{2}} V_{us} V^{*}_{cs}
  C_7 \frac{e}{4\pi^2} m_{c} \bar u \sigma_{\mu \nu}
  P_{R}c F^{\mu \nu}\,,
 \ed
where $C_7=C^{SM}_7 + C^{NP}_7$ and $C^{SM}_7\approx -(0.007 + i
0.02)=0.021e^{i\delta_s}$ with $\delta_s=70.7^{\circ}$ \cite{GHMW}
being the strong phase induced by the two-loop QCD corrections.
In the extension of the SM by including
a weak singlet
particle, the flavor mixing matrix in the charged current is not unitary
and the $Z$-mediated FCNC at tree level is generated as well.
For $c\to u\ga$, besides the QED-penguin diagrams induced by the $W$-boson displayed in
Figs.~\ref{fig:WZ}a and \ref{fig:WZ}b,
%%%%%%%%%%%%%%%%%%%%%%%%%%%%%%%%%%%%%%%%%%%%%%%%%%%%%%%%%%%%%%%%%%%%%%%%%
%%%%%%%%%%%%%%%%%%%%%%%%%%%%%%%%%%%%%%%%%%%%%%%%%%%%%%%%%%%%%%%%%%%%%%%%%
the $Z$-mediated QED-penguin one
%,
in Fig.~\ref{fig:WZ}c will also give contributions.
 We note that the contributions from $W_H$
and $Z_H$ can be ignored
as $m^2_{W}/m^{2}_{W_H}$ and $m^2_{Z}/m^{2}_{Z_H}$ are much
less than one.

\begin{figure}[htbp]
\includegraphics*[width=4.5in]{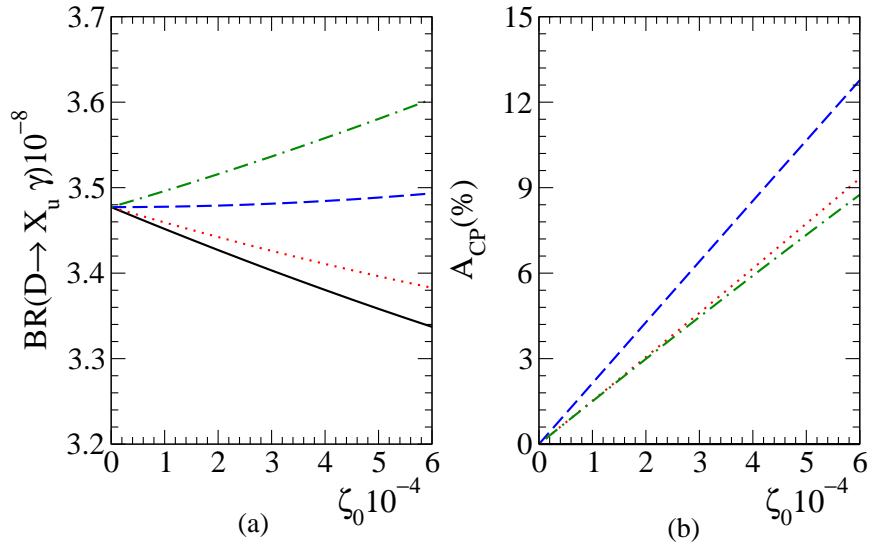}
\caption{BR (in units of $10^{-8}$) and CPA (in units of $10^{-2}$)
for $D\to X_u \ga$ as functions of $\zeta_0$, where the solid,
dotted, dashed and dash-dotted lines represent the CP violating
phase at $\theta=0$, $45^{\circ}$, $90^{\circ}$ and $135^{\circ}$,
respectively.}
 \label{fig:brcp}
\end{figure}

At the first sight, due to the light quarks in the loops,
the contributions from Figs.~\ref{fig:WZ}a and \ref{fig:WZ}b
could be negligible. However, due to the  non-unitarity of
$(VV^{\dagger})_{uc}=V_{14} V^{*}_{24}\neq 0$, even in the limits
of
$m_{d,\,s,\,b}\to 0$, the contributions from the mass independent terms
do not vanish anymore and can be sizable.
In terms of unitary gauge \cite{CCK_PRD61},
we obtain
 \be
  C^{W}_7 &=& \frac{11}{36} \frac{(V V^{\dagger})_{12}}{ V_{us} V^*_{cs}}
  = \frac{11}{36} \frac{V_{14} V^{*}_{24}}{V_{us} V^*_{cs} } \,.
 \ed
Furthermore, if we set $m_u\approx m_c =0$, the contributions from
Fig.~\ref{fig:WZ}c are given by
\be
C^{Z}_7&=&(f^{Z}_{u} + f^{Z}_{c} +
f^{Z}_{T})/V_{us}V^{*}_{cs}\,,\non\\
  f^{Z}_{c}+ f^{Z}_{u} &\approx &  \frac{1}{2} e_{u}
   \left\{
 \left( \frac{1}{2}
 -e_{u} \sin^2\theta_W \right)
 \left[4 \xi_0(0) -6 \xi_1(0) +2 \xi_{2}(0) \right] \right.\non \\
 &&  \left. + e_u \sin^2\theta_W \left[ 4 \xi_0(0) -4 \xi_{1}(0)
 \right]\right\}\,,\non\\
  f^{Z}_{T}&=& \frac{1}{4} e_u
  \left[2\xi_{0}(y_T)-3\xi_1(y_T)
  +\xi_2(y_T)\right]\,,
 \ed
where the functions $\xi_{n}(x)$ are defined by
\be
\xi_{n}(x)\equiv\int^{1}_{0}{z^{n+1}dz\over 1+(x-1)z}
\ed
 and  $y_T=m_T/m_Z$.
 Numerically, the total contribution in Fig.~\ref{fig:WZ} is
 \be
 C_7= C^W_7 + C^Z_7\approx  \frac{0.53}{V^*_{cs} V_{us}} V_{14} V^{*}_{24}\,.
 \ed
If we regard $V_{14} V^{*}_{24}$ as an unknown complex parameter, $i.e.$
$V_{14} V^{*}_{24}= \zeta_0 e^{i\theta}$ with
$\theta$ being the CP violating phase, one can study the decay BR and
direct CP asymmetry (CPA) of $D\to X_u\ga$ defined by
 \be
  {\rm BR}(D\to X_u\ga)&=&\frac{6 \alpha_{\rm em} |C_7|^2}{\pi|V_{cd}|^2}
  {\rm BR}(D\to X_d e \bar \nu_{e})\,, \non\\
A_{CP}&=& \frac{{\rm BR}(\bar c\to \bar u \ga)- {\rm BR}(c\to  u \ga)}{{\rm BR}(\bar
c\to \bar u \ga) + {\rm BR}(c\to  u \ga)}\,, \label{eq:brcp}
 \ed
 as functions of $\zeta_0$ and $\theta$.
 In Fig.~\ref{fig:brcp},
the BR and CPA as  functions of $\zeta_0$ are presented,
 where the solid, dotted, dashed and dash-dotted
lines represent the CP violating phase at $\theta=0$, $45^{\circ}$,
$90^{\circ}$ and $135^{\circ}$, respectively.
%%%%%%%%%%%%%%%%%%%%%%%%%%%%%%%%%%%%%%%%%%%%%%%%%%%%%%%%%%%%%%%%%%%%%%%%%
%%%%%%%%%%%%%%%%%%%%%%%%%%%%%%%%%%%%%%%%%%%%%%%%%%%%%%%%%%%%%%%%%%%%%%%%%
 From these results, it is interesting to see
 that
${\rm BR}(D\to X_u\ga)$ is insensitive to the new physics effects, whereas
the direct CPA  could be as large as $O(10\%)$.
Explicitly, if we take $\theta=90^{\circ}$ and
$\zeta_{0}=1.5\times 10^{-4}$,
the CPA is about $ 3\%$.
Note that this CPA vanishes in the SM.

\subsection{$D\to X_u\ell\bar{\ell}$ and $D^{0}\to \ell^{+} \ell^{-}$
decays}

Because the current experimental measurements in $K$ and $B_q$
decays are all consistent with the SM predictions, it is inevitable that
if we want to observe any deviations from the SM, we have to wait
for precision measurements for $K$ and $B_q$. SuperB factories or
LHCb could provide a hope. However, the situation in $D$ physics
is straightforward. As stated before, unlike $K$ and $B_q$
systems, due to no heavy quark enhancement in the $D$ system,
the rare $D$-meson decays,
such as $D\to X_u \ell \bar \ell$ ($\ell=$ e, $\mu$,
$\nu$), are always suppressed. Even by considering the
long-distance effects, the related decays, such as $D\to \mu^{+}
\mu^{-}$ and $D\to X_u\nu \bar \nu$, get small corrections to
the SD predictions on the BRs \cite{BGHP_PRD66}.
Therefore, these rare decays definitely
could be good candidates to probe the new physics effects. Since the values in the SM
are hardly reachable
at $D$ factories \cite{bes3},
if any exotic event is found, it must be a strong
evidence for new physics. In the following analysis, we are
going to discuss the implication of the Littlest Higgs model on the rare $D$
decays involving di-leptons.

To study these decays, we first write the effective Hamiltonian
for $c\to u \ell^{+} \ell^{-}$ ($\ell=e,\, \mu$) as
 \be
 {\cal H}(c\to u \ell^{+} \ell^{-})&=& -\frac{G_F\alpha_{\rm
  em}}{\sqrt{2}\pi}V^{*}_{cs} V_{us}
  \left[C^{\rm eff}_9 O_9 + C_7 O_{7} + C _{10} O_{10}\right]\,,
  \label{eq:hcull}\\
    O_{7}&=& -\frac{2m_c}{q^2} \bar u i \sigma_{\mu\nu} q^{\nu} P_{R} c
\bar\ell \ga^{\mu} \ell \,,
\non \\
  O_{9} &=& \bar u\ga_{\mu} P_L c\, \bar\ell \ga^{\mu} \ell \,,
\non \\
  O_{10} &=& \bar u\ga_{\mu} P_L c\, \bar\ell \ga^{\mu} \ga_5 \ell \,,
  \label{eq:cull}
  \end{eqnarray}
where the effective Wilson coefficients are given by
 \be
 C^{\rm eff }_{9} &=&
 \frac{2\pi}{\alpha_{\rm em}}  \frac{(VV^{\dagger})_{14}}{V^{*}_{cs} V_{us}} c^{\ell}_{V}
 +\left(h(z_s, s)-h(z_d, s)\right)\left(C_2(m_c)+3C_1(m_c) \right)
   \,, \non \\
  C_{10} &=& - \frac{2\pi}{\alpha_{\rm em}}  \frac{(VV^{\dagger})_{14}}{V^{*}_{cs} V_{us}}
  c^{\ell}_{A}\,,
 \ed
 with
 $s=q^2/m^2_c$, $z_i=m_i/m_c$,
$c^{\ell}_{V}=-1/2+2\sin^2\theta_W$, $c^{\ell}_{A}=-1/2$ and
 \be
h(z, s)&=& -\frac{4}{9} \ln z+ \frac{4}{27}+ \frac{2}{9} x
-\frac{1}{9} \left( 2+ x \right)\sqrt{|1-x|}
 \non \\
 &\times &
   \left\{
  \begin{array}{c}
    \ln \left|\frac{\sqrt{1-x}+1}{\sqrt{1-x}-1} \right|-i\, \pi, \
    {\rm for}\; x\equiv 4z^2/ s<1 \, , \\
    2\, {\rm arctan}\frac{1}{\sqrt{x-1}},\   {\rm for}\; x \equiv 4z^2/ s>1   \, \\
  \end{array}\right.
  \,.
 \ed
 Here,
we have neglected the small contributions from the penguin and box
diagrams. We note that in the SM, the SD contributions are mainly from
the term with $h(z, s)$, induced by the insertion of
$O_2=\bar u_{L}\ga_{\mu} q_L \bar q_{L} \ga^{\mu} c_{L}$ and mixing
with $O_9$ at one-loop level \cite{BGHP_PRD66,FSZ}.
We note that the
resonant decays of $D\to X_u V\to X_u \ell^{+} \ell^{-}$ $(V=\phi$,
$\rho$, $\omega)$ would have large corrections to $c\to
u\ell^{+} \ell^{-}$ at the resonant regions.
However,  in this paper
we do not discuss
these contributions as we only
concentrate on the SD contributions.
Moreover, these resonance contributions can be
removed by imposing proper cuts in the phase space in dedicated searches. %

\begin{figure}[htbp]
\includegraphics*[width=3.in]{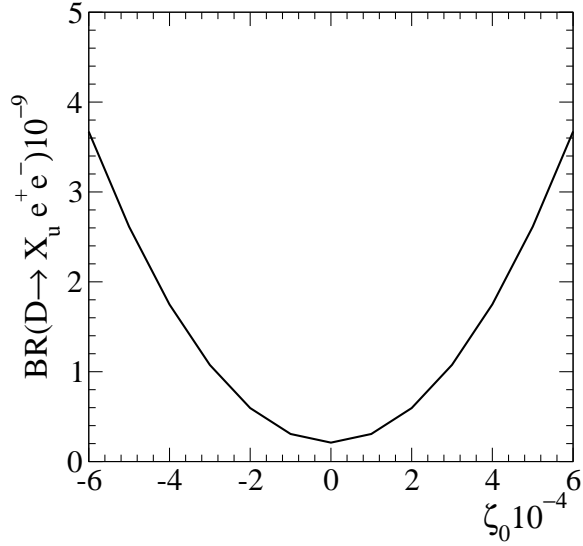}
\caption{BR(in units of $10^{-9}$) for $D\to X_u e^{+} e^{-}$ as a
function of $\zeta_0$. }
 \label{fig:cull}
\end{figure}

 From Eq.~(\ref{eq:hcull}), the decay rate for $D\to X_u \ell^{+}
\ell^{-}$ as a function of the invariant mass $s=q^2/m^2_c$ can be
found to be
 \be
\frac{d\Gamma}{  ds}&=& \frac{G^2_F m^5_c \alpha^2_{\rm
em}}{768\pi^5} |V_{us} V^{*}_{cs}|^2(1-s)^2 R(s)\,,\non\\
R(s)&=& \left( |C^{\rm eff}_{9}|^2+ |C_{10}|^2\right) (1+2 s)+12
Re(C^{*}_{7} C^{\rm eff}_9) + 4\left( 1+\frac{2}{s}\right)|C_7|^2\,.
\label{eq:ratecull}
 \ed
In addition, by utilizing the lepton angular distribution, we can
also study the forward-backward asymmetry (FBA),  given by
 \be
 \frac{d A_{FB}}{ ds}&=& {\int ^{1}_{-1} d\cos\theta d\Gamma/
 dsd\cos\theta
 \ {\rm sgn(\cos\theta)}\over \int^{1}_{-1} d\cos\theta
 d\Gamma/dsd\cos\theta}\,, \non \\
  &=&  -3 \frac{s}{R(s)} {\mathrm Re} \left[ \left(C^{\rm
eff}_9 + \frac{2}{s} C_7\right) C^{*}_{10}\right]\,,
  \ed
where
 $\theta$ is the angle of $\ell^{+}$ related to
the momentum of the $D$ meson in the $\ell^{+} \ell^{-}$ invariant mass
frame. Since $C_{10}$ is small in the SM, $A_{FB}$ is negligible. With
$m_c=1.4$ GeV and the mixing parameter in Eq. (\ref{eq:limit}), we
get
  \be
  {\rm BR}(D\to X_u e^+ e^-)&=&  (4.18\pm 0.91) \times 10^{-10}\,,\non \\
  {\rm BR}(D\to X_u \mu^{+} \mu^{-})&=& (2.51\pm 0.86) \times 10^{-10}\,,
 \ed
comparing with
 the SM predictions of $
  {\rm BR}(D\to X_u e^+ e^-)_{SM}= 2.1 \times 10^{-10}$ and
  ${\rm BR}(D\to X_u \mu^{+} \mu^{-})_{SM}= 0.5 \times 10^{-10}$, respectively.
 Clearly,
if some cancellation occurs between new physics and SM contributions
in the $D-\bar D$ mixing, a larger value of $\zeta_0$ can be
allowed.
In Fig.~\ref{fig:cull},
we show the tendency of the decay as a function of $\zeta_0$, where
the negative horizontal values correspond to -$\zeta_0$.
%%%%%%%%%%%%%%%%%%%%%%%%%%%%%%%%%%%%%%%%%%%%%%%%%%%%%%%%%%%%%%%%%%%%%%%%%
%%%%%%%%%%%%%%%%%%%%%%%%%%%%%%%%%%%%%%%%%%%%%%%%%%%%%%%%%%%%%%%%%%%%%%%%%
%%%%%%%%%%%%%%%%%%%%%%%%%%%%%%%%%%%%%%%%%%%%%%%%%%%%%%%%%%%%%%%%%%%%%%%%%
\begin{figure}[htbp]
\includegraphics*[width=4.5in]{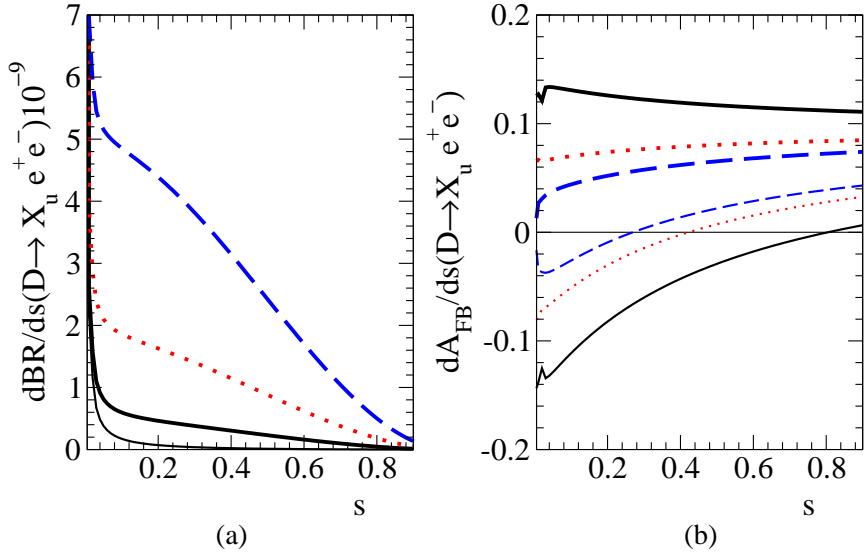}
\caption{(a)[(b)] Differential BR (in units of $10^{-9}$) [FBA] for
$D\to X_u e^{+} e^{-}$ as a function of $s$, where the thick solid,
dotted and dashed lines correspond to $\zeta_0=1.5$, $3.0$ and
$5.0$, while the thin ones denote the cases for $-\zeta_0$ except
$\zeta_{0}=0$ for the thin solid line in (a).  }
 \label{fig:difcull}
\end{figure}
%%%%%%%%%%%%%%%%%%%%%%%%%%%%%%%%%%%%%%%%%%%%%%%%%%%%%%%%%%%%%%%%%%%%%%%%%
In addition, we present the differential decay BR [FBA] of $D\to X_u
e^{+} e^{-}$ as a function of $s=q^2/m^2_c$ in
Fig.~\ref{fig:difcull}a [b], where the thick solid, dotted and
dashed lines correspond to $\zeta_0=1.5$, $3.0$ and $5.0$, while the
thin ones denote the cases for $- \zeta_0$ except $\zeta_{0}=0$ for
the thin solid line in Fig.~\ref{fig:difcull}a.
%{
 From Fig.~\ref{fig:difcull}b, we see
that the FBA is only at percent level. In the Littlest Higgs model,
this is because the $Z$ coupling to the charged lepton
$c^{\ell}_{V}=-1/2+2\sin^2\theta_W$ appearing in $C^{\rm eff}_{9}$
is much smaller than one. This is quite different from that in $b\to
s\ell^{+} \ell^{-}$ where the dominant effect in the SM for the FBA
is from the box and QED-penguin diagrams.

\begin{figure}[htbp]
\includegraphics*[width=4.5in]{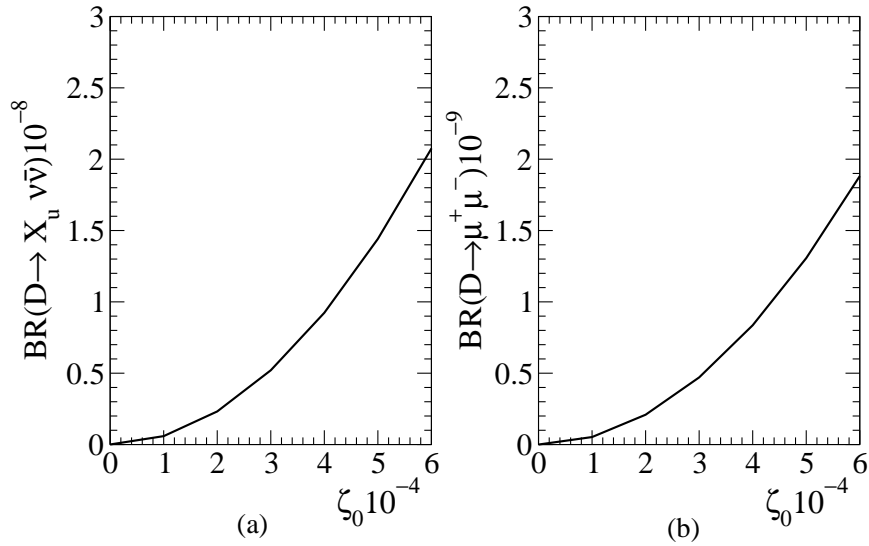}
\caption{(a) BR (in units of $10^{-8})$ for $D\to X_u \nu \bar\nu$ and
(b) BR (in units of $10^{-9}$) for $D\to \mu{+} \mu^{-}$. }
 \label{fig:cununu}
\end{figure}

Next, we discuss the decay of $D\to X_u \nu \bar \nu$. In the SM,
the BR for $D\to X_u \nu \bar \nu$
is estimated to be $O(10^{-16})-O(10^{-15})$ \cite{BGHP_PRD66},
which is vanishing small.
In the Littlest Higgs model, by
taking $C_7=0$, $C^{\rm eff}_9=-C_{10}= -\pi
(VV^{\dagger})_{14}/(\alpha_{\rm em} V_{us} V^{*}_{cs})$, the
effective Hamiltonian in Eq.~(\ref{eq:hcull}) can be directly
applied to $c\to u \nu \bar\nu$.
The decay rate
for $D\to X_u \nu \bar\nu$ as a function of $s=q^2/m^2_c$ can be
obtained as
 \be
 \frac{d\Gamma}{  ds}&=& 3 \frac{G^2_F m^5_c }{768\pi^5}
 (1-s)^2(1+2s)
\left( 2 \pi^2 |(VV^{\dagger})_{12}|^2\right)\,,
 \ed
where the factor of $3$ stands for the neutrino species. With
$\zeta_0=1.5 \times 10^{-4}$, we get BR$(D\to X_u \nu \bar\nu)=1.31
\times 10^{-9}$. However, if we relax the constraint on
$V_{14}V^{\dagger}_{24}$, the BR as a function of $\zeta_{0}$ is
shown in Fig.~\ref{fig:cununu}a.
For a larger value of $\zeta_{0}$,
BR$(D\to X_u \nu \bar\nu)$ could be as large as
$O(10^{-8})$.
%%%%%%%%%%%%%%%%%%%%%%%%%%%%%%%%%%%%%%%%%%%%%%%%%%%%%%%%%%%%%%%%%%%%%%%%%
%%%%%%%%%%%%%%%%%%%%%%%%%%%%%%%%%%%%%%%%%%%%%%%%%%%%%%%%%%%%%%%%%%%%%%%%%

Finally, we study the decays of $D\to \ell^{+} \ell^{-}$. It has
been known that, in the SM, the SD contributions to $D\to \mu^{+}
\mu^{-}$ are $O(10^{-18})$, while the LD ones are $O(10^{-13})$
\cite{BGHP_PRD66}.
It is clear that any signal to be observed at the
sensitivity of the proposed detector, such as BESIII, will indicate
new physics effects. Since the  effective interactions at
quark level are the same as those in Eq.~(\ref{eq:hcull}),
one finds that
 \be
  {\rm BR}(D\to \ell^{+} \ell^{-})&=& \frac{G^2_F
}{16\pi^2}  \tau_D m_D m^2_{\ell} f^2_D
\sqrt{1-\frac{4m^2_{\ell}}{m^2_D}} |\pi V_{14}V^{*}_{24}|^2\,.
 \ed
Here we have used equation of motion for the charged lepton so that
$\bar \ell\slash{\!\!\!{p}}_{D} \ell =0$. We also note that operators
$O_{7,9}$ make no contributions. With
$|V_{14}V^{*}_{24}|=\zeta_0=1.5 \times 10^{-4}$, the predicted BR
for $D\to \mu^{+} \mu^{-}$ is $1.17 \times 10^{-10}$. In
Fig.~\ref{fig:cununu}b,
 we present the BR as a
function
of $\zeta_0$.
We see that  BR$(D\to \mu^{+} \mu^{-})$
in the Littlest Higgs model could be as large as $O(10^{-9})$. \\

\section{Conclusions}
\label{sec:Concl}

We have studied the $D-\bar D$ mixing and rare $D$ decays in the Littlest Higgs model. In the model,
as the new weak singlet vector-like quark $T$ with the electric charge
of $2/3$ is introduced to cancel the quadratic divergence
induced by the top-quark,
 the standard unitary $3\times 3$
CKM matrix is extended to a non-unitary
$4\times 3$ matrix in the quark charged currents and
$Z$-mediated flavor changing neutral currents
are generated at tree level. We have shown that
the effects on $|\Delta C|=2$ and $|\Delta C|=1$ processes are all related to
$V_{14}V^{*}_{24}$ in Eq.~(\ref{eq:nonunitary}).

To avoid the scenario adopted by Ref.~\cite{Lee}, in which
$\l_0\sim \l_{33}\gg \l_{ij}$ was assumed, we choose the basis such that
the effective mass matrix for $u_1$, $c_2$ and $t_3$ is diagonal,
while the corresponding masses $m_1$, $m_2$ and $m_3$ are free
parameters and can be as large as the weak scale $v$. Since the global
symmetry breaking scale $f$ is larger than $v$, the
mixing matrix relating physical and unphysical states could be
extracted by taking the leading perturbative expansion. Accordingly, by
using the approximation of $m_u\approx m_c \approx 0$, the explicit
expressions for $V_{14}$ and $V_{24}$ have been obtained.
In terms of the data for $V_{ub}$ and $V_{cb}$, we
have found that the natural value for $\zeta_{0} \equiv|V_{14}V^{*}_{24}|$ is $O(10^{-4})$, which agrees with the observed parameter in the $D-\bar{D}$ mixing but
it is one order of
magnitude larger than that in Ref.~\cite{Lee}.

For the rare $D$ decays,
due to the non-unitarity effects in the model,
BR$(D\to X_u \ell^{+} \ell^{-})$, BR$(D\to X_u\nu \bar \nu)$ and
BR$(D\to \mu^{+} \mu^{-})$ could be
enhanced to be $O(10^{-9})$, $O(10^{-8})$ and $O(10^{-9})$,
respectively,
which could  marginally reach the sensitivity proposed by BESIII
\cite{bes3}.\\

 \noindent {\bf Acknowledgments}

This work is supported in part by the National Science Council of
R.O.C. under Grant \#s: NSC-95-2112-M-006-013-MY2,
 NSC-95-2112-M-007-001, NSC-95-2112-M-007-059-MY3 and
 NSC-96-2918-I-007-010.

\end{document}